# Observation of Dirac node formation and mass acquisition in a topological crystalline insulator

Yoshinori Okada[1,2], Maksym Serbyn*[3], Hsin Lin*[4], Daniel Walkup[1], Wenwen Zhou[1], Chetan Dhital[1], Madhab Neupane[5], Suyang Xu[5] Yung Jui Wang[4], R. Sankar[6], Fangcheng Chou[6], Arun Bansil[4], M. Zahid Hasan[5], Stephen D. Wilson[1], Liang Fu[3] and Vidya Madhavan[1]

In the recently discovered topological crystalline insulators (TCIs), topology and crystal symmetry intertwine to create surface states with a unique set of characteristics. Among the theoretical predictions for TCIs is the possibility of imparting mass to the massless Dirac fermions by breaking crystal symmetry, as well as a Lifshitz transition with a change of Fermi surface topology. Here we report high resolution scanning tunneling microscopy studies of a TCI, $Pb_{1-x}Sn_xSe$. We demonstrate the formation of zero mass Dirac fermions protected by crystal symmetry and the mechanism of mass generation via symmetry breaking, which constitute the defining characteristics of TCIs. In addition, we show two distinct regimes of fermiology separated by a Van-Hove singularity at the Lifshitz transition point. Our work paves the way for engineering the Dirac band gap and realizing interaction-driven topological quantum phenomena in TCIs.

[1]Department of Physics, Boston College, Chestnut Hill, Massachusetts 02467, USA

[2] WPI-AIMR, Tohoku University, Sendai, 980-8577, Japan

[3] Department of Physics, Massachusetts Institute of Technology, Cambridge, MA 02139,USA

[4] Department of Physics, Northeastern University, Boston, Massachusetts 02115, USA

[5]Joseph Henry Laboratory, Department of Physics, Princeton University, Princeton, New Jersey 08544, USA

[6]Center for Condensed Matter Sciences, National Taiwan University, Taipei 10617, Taiwan

*These authors contributed equally to this work

Topological crystalline insulators (*1,2,3,4,5*) (TCIs) in IV-VI semiconductors possess a unique type of electronic topology protected by crystalline symmetry, which gives rise to surface states with an extraordinary band dispersion and spin texture. In particular, the (001) surface harbors two generations of Dirac cones at different energies. The band structure of (001) surface states can be visualized by starting from a pair of Dirac points located at the edge of the surface Brillouin zone ($\bar{X}$ point in Fig.1), with the Dirac point energies close to the conduction and valence band edge respectively (*6*). The hole-branch of the upper Dirac cone and the electron-branch of the lower Dirac cone coexist inside the band gap. The hybridization between these two branches leads to an avoided crossing in all directions except along the $\Gamma\bar{X}$ line where a band crossing is dictated by electronic topology of TCI and protected by the (110) mirror symmetry. Such a band reconstruction generates a new pair of low-energy Dirac nodes displaced symmetrically away from each $\bar{X}$ point as shown in Fig. 1.

The distinctive band structure of the TCI surface states has two important consequences. First, the formation of low-energy Dirac nodes is accompanied by a Lifshitz transition (*7*): as the Fermi energy moves deep into the band gap, the Fermi surface changes from concentric pockets of opposite carrier types centered at $\bar{X}$ into a pair of disconnected pockets across the Brillouin zone boundary as depicted in Fig.1 and Fig.3. The change of Fermi surface topology occurs via a saddle point in the surface band structure, which has been predicted to lead to a divergence in the electronic density of states known as Van Hove singularity (VHS) at the saddle-point energy (*6*). While a VHS generally enhances interaction effects and drives electronic instabilities to magnetism, charge density waves or superconductivity (*8,9,10,11*), VHSs are particularly interesting in two-dimensional TCI surface states where interaction effects in combination with band topology may result in novel correlated topological states (*12,13*). Second, the fundamental role of crystalline mirror symmetry in protecting the Dirac nodes implies an ideal tunability of electronic properties of Dirac fermions. One intriguing prediction is that when mirror symmetry is broken---either spontaneously or by external perturbations, the TCI surface states acquire a gap thereby creating massive Dirac fermions with unique topological characteristics (*2*) and providing an exciting new avenue to control the properties of Dirac materials by strain or doping. Such Dirac mass generation due to symmetry breaking has not yet been experimentally observed.

We used a low temperature scanning tunneling microscope (STM) to track the dispersion and the density of states of $Pb_{1-x}Sn_xSe$ with a nominal doping of x=0.5 (see supplementary information for determination of actual doping level at x=0.34) which lies in the topological crystalline insulator phase (*5*). By using Landau level (LL) spectroscopy at different magnetic fields, we have mapped out the detailed band dispersion and the density of states of TCI surface states with meV resolution and are able to identify features that were inaccessible in previous angle-resolved photoemission spectroscopy (ARPES) experiments on TCIs. By comparing our data with theory, we accurately determine, for the first time the key parameters for the TCI surface band structure. In particular, our data clearly show the presence of VHSs at a critical energy separating two distinct types of fermiology, as well as an unexpected feature that we identify as the unique signature of mass generation of two-dimensional Dirac fermions.

Single crystal $Pb_{1-x}Sn_xSe$ samples were cleaved in UHV at approximately 80K before being inserted into the STM and all measurements were obtained at 4K. Although $Pb_{1-x}Sn_xSe$ has a 3D crystal structure (Fig. 1A), it cleaves naturally along the [001] plane resulting in the Pb/Sn-Se surface shown in Fig. 2A. STM topography reveals a square lattice whose inter-atom spacing of 4.32Å indicates a preferential imaging of either the Pb/Sn atoms or Se atoms at a particular bias voltage. The Se lattice is expected to be homogeneous while the Pb/Sn lattice should reveal a contrast corresponding to the presence of either Pb or Sn atoms. Since the positive bias topographies reveal a significant percentage of defects centered at lattice sites, we conclude that the atoms being imaged at these bias voltages are Pb or Sn. In addition to the 1X1 lattice we also observe an incoherent $\sqrt{2}$ superstructure, which most likely has an electronic origin arising from the nesting of the Dirac cones connected by the $\sqrt{2}$ scattering vector. The symmetry of the $\sqrt{2}$ structure is such that none of the critical symmetries for topological phase formation is broken and it is therefore not expected to have a dramatic effect on the surface band structure.

The overall density of states in this TCI can be obtained by measuring *dI/dV* spectra. Fig. 1C shows a plot of a typical *dI/dV(eV)* spectrum in this material, where the spectrum shows a V-shaped density of states with a minimum at ~-80meV. In addition, two peaks are observed on either side of the minimum at ~ -40 meV and ~ -120meV, respectively. By comparing the density of states to the schematic band structure, we tentatively assign the minimum at -80meV to the Dirac point labeled $E_{DP1}$. The symmetric peaks on either side of the Dirac point are more difficult to identify since they may arise from either the surface state or the bulk bands, and further data on the band structure is needed to distinguish their origin. In the following we develop a framework which combines our magnetic field dependent STM data with a theoretical model to establish the surface state dispersion.

In order to present spectra that are representative of the material it is important to understand the level of inhomogeneity in these samples. To do this, we compare spectra obtained at various spatial locations. As shown by the spectral linecut in Fig. 2E, the surface of $Pb_{1-x}Sn_xSe$ shows a remarkable degree of spectral homogeneity over at least 300Å, despite the presence of randomly distributed Sn atoms within the Pb lattice. This is in contrast with the highly inhomogeneous nature of graphene as well as doped topological insulators such as $Bi_2Se_3$/$Bi_2Te_3$ (*14*,*15*). This important feature makes $Pb_{1-x}Sn_xSe$ a much more stable host for topological surface states, and allows one to truly access physics at the Dirac point with a variety of experimental probes. Taking advantage of this homogeneity, we henceforth use linecut-averaged spectra along the line shown in Fig. 2A for our analysis. The evolution of line-cut averaged spectra with magnetic field is shown Fig. 3A, where the Landau level peaks are clearly visible. For normal 2D bands with linear or quadratic dispersion, the semi-classical approximation is applicable and a plot of the LL peak position as a function of *nB* (where n is the LL index and B is the magnetic field strength) can be used to obtain information on the dispersion (*15,16*). While the surface state evolution with energy in TCIs is predicted to be complex, we find that LL spectroscopy nevertheless provides a very effective means of experimentally accessing the resulting dispersion with high energy resolution over a wide energy range.

Comparing the zero field spectrum with the spectra at higher fields as shown in Fig. 3A, we find that for non-zero magnetic fields, a peak located precisely at the density of states minimum (at ~-80meV) emerges. It is non-dispersive, i.e. its position does not change with magnetic field which confirms its origin as the $0^{th}$ LL located at the Dirac point, $E_{DP1}$. In addition, we find other non-dispersing peaks which have been labeled $E_-^*$, $E_+^*$ and $E_{DP2}$ in Fig. 3A. The origin of these peaks will be discussed later in the paper. To understand the electronic structure and characteristic energy scales, as a first step we analyze the LL data within a semi-classical picture. This requires us to index the dispersing LL peaks, which is a non-trivial task in this material. For indexing, we use a procedure that utilizes the scaling behavior of the LL peaks with magnetic field. From the theory (*6*) we know that at energies away from the saddle points the band structure is characteristic of Dirac fermions with approximately linear dispersion. Therefore, we expect the peak energies to collapse to one curve as a function of $\sqrt{nB}$. This scaling behavior provides a strong constraint on the peak index assignments for the higher and lower LLs where this is most applicable. The resulting plot of the LL energy with $\sqrt{nB}$ is shown in Fig. 3C. Comparing this to the zero field spectrum we find that a sharp discontinuity in the LL plot occurs at the energy scale of the peaks in the zero field spectrum ~40meV on either side of $E_{DP1}$. Putting this information together with the schematic dispersion in Fig 1, we assign the zero field *dI/dV* peaks to VHSs, which is predicted to arise from the saddle points in surface state dispersion. This is further confirmed by comparing the theoretically predicted Femi surface area with our data as shown in Fig. 3D (for details refer to supplementary information) and verified again later in this paper when we calculate the dispersion using our LL data.

Our observation of VHSs in a topologically non-trivial material has important ramifications. The primary challenge in realizing novel correlated phases in Dirac materials is typically the low density of electronic states of the surface state electrons and the correspondingly weak electron-electron interactions. The existence of a VHS close to the Fermi energy therefore opens the exciting possibility of achieving correlated states in this Dirac material. While VHSs have previously been experimentally observed in the Dirac material graphene (*17*), the topological nature of the band structure in TCIs creates a potential for novel phases that goes beyond what can be expected in graphene. This is because symmetry breaking interactions modifying existing topologically protected electronic states have the potential of generating unique states such as helical domain wall states (*2*), or Majorana fermion modes (*18*).

We now discuss the intriguing appearance of the magnetic field induced peaks labeled $E_-^*$, $E_+^*$

($\sim\pm$10meV from $E_{DP1}$) in Fig. 3A. The immediate striking observation is that the peak energies do not change with magnetic field strength. This rules out g-factor effects or a Zeeman term. Furthermore, the appearance of the additional peaks is restricted to the vicinity of the Dirac point $E_{D1}$, while all the other LL peaks (except $E_{DP2}$, which we discuss later) are accounted for by the surface state dispersion. The only explanation for these peaks is that the Dirac node at zero-field is gapped out by the acquisition of a small mass term. A unique property of the resulting massive two-dimensional Dirac fermions is the appearance of a non-dispersing n=0 LL pinned at the energy of the Dirac mass, which as we show below is exactly what is observed in our data.

The coexistence of the peaks at $E_-^*$ and $E_+^*$ with the zero–mode LL peak at the gapless Dirac point places strong constraints on the origin of the Dirac mass. In TCIs, such a scenario is only realized when mirror symmetry is broken. Since there are two mirror planes within the surface BZ, it is possible to selectively break mirror symmetry reflection with respect to one mirror plane such as (110), leaving the other mirror plane unaffected. This can be achieved for example, by a rhombohedral or orthorhombic distortion of the crystal structure with a displacement of atoms at the surface as shown in Fig. 4E, which is a common distortion in this class of materials (*19,20,21*). One way to confirm this scenario is to calculate the Landau level spectra theoretically while taking the broken symmetry into account, which can then be compared to the experimental data. To do this we modify the recently developed **k.p** Hamiltonian for TCIs (*6*) by adding a new symmetry breaking mass term which results in the surface state band structure shown in the schematic in Fig. 4E (see supplementary information and *(22)* for more details on the theory). Using this model the LL spectrum is calculated as a function of magnetic field and presented as a fan diagram in Fig. 4. Comparing the LL fan-diagram without (Fig. 4B) and with (Fig. 4C) the symmetry breaking term, we see that the mass term (with an appropriately chosen mass of $\sim$11 meV) partially splits four-fold degenerate $0^{th}$ LL, while the LL spectra at higher energies are not substantially affected.

The details of our data fit extraordinarily well with our theoretical model for coexisting massive and massless Dirac fermions. The symmetry breaking term gaps out two of the four nodes but the sign of the mass term for the two gapped nodes is necessarily opposite, as dictated by time reversal symmetry (*2*). As a consequence, the n=0 LLs of these two massive Dirac fermions appear on the upper and lower branch of the spectrum respectively, each of which is non-degenerate. This beautifully explains our observation of two symmetric peaks near the Dirac point. All the other Landau levels of each gapped Dirac node are expected to be particle-hole symmetric, thereby resulting in a two-fold valley degeneracy, only weakly affected by the symmetry breaking. We note that while the energies of the Landau levels above the Dirac point $E_{DP1}$ fit very well with our theoretical particle-hole symmetric dispersion, the LLs below $E_{DP1}$ show small deviations. This asymmetry is however not discussed further in this paper. By fitting the theoretical LLs to our data we calculate the surface state dispersion as shown in Fig. S3, which reveals that the saddle point energy is $\sim$40 meV from the Dirac point, once again consistent with our previous identification of the zero-field dI/dV peaks with VHSs. Remarkably, this analysis also provides an independent identification of the non-dispersing feature labeled $E_{DP2}$. Directly after the Dirac cones merge, a secondary Dirac cone appears at $\bar{X}$, as shown schematically in Fig. 1D and our calculations in Fig. S3. Theoretically this cone also results in LLs. While the most of the LL peaks associated with this secondary cone are not experimentally accessible at our fields, the non-dispersing peak in the experimental spectra at $E_{DP2}$ lies at the correct energy to be identified as the Dirac point associated with the secondary Dirac cone (Fig. 1D).

In summary, our data demonstrate an enhanced, nearly singular, density of states inside in the bulk band gap and tied to the surface state spectrum near the Dirac point. We show that this enhancement stems from nearby Van Hove singularities unique to the surface state dispersion of this new category of topological materials. Our results also reveal a broken mirror symmetry in this TCI leading to the gapping of select Dirac nodes. Symmetry breaking interactions modifying existing topologically protected electronic states have the potential of generating new phases such as helical quantum wire states which can emerge at domain boundaries (without an applied magnetic field) and a plethora of other novel electronic phenomena. The band structure enabled by the coexistence of both massive and massless Dirac fermions in the same surface spectrum as well as the enhanced density of states in close proximity to the Dirac point demonstrate that $Pb_{1-x}Sn_xSe$ constitutes a realistic, tunable, platform the for exploring novel topological states emergent via coupling to symmetry breaking interactions.

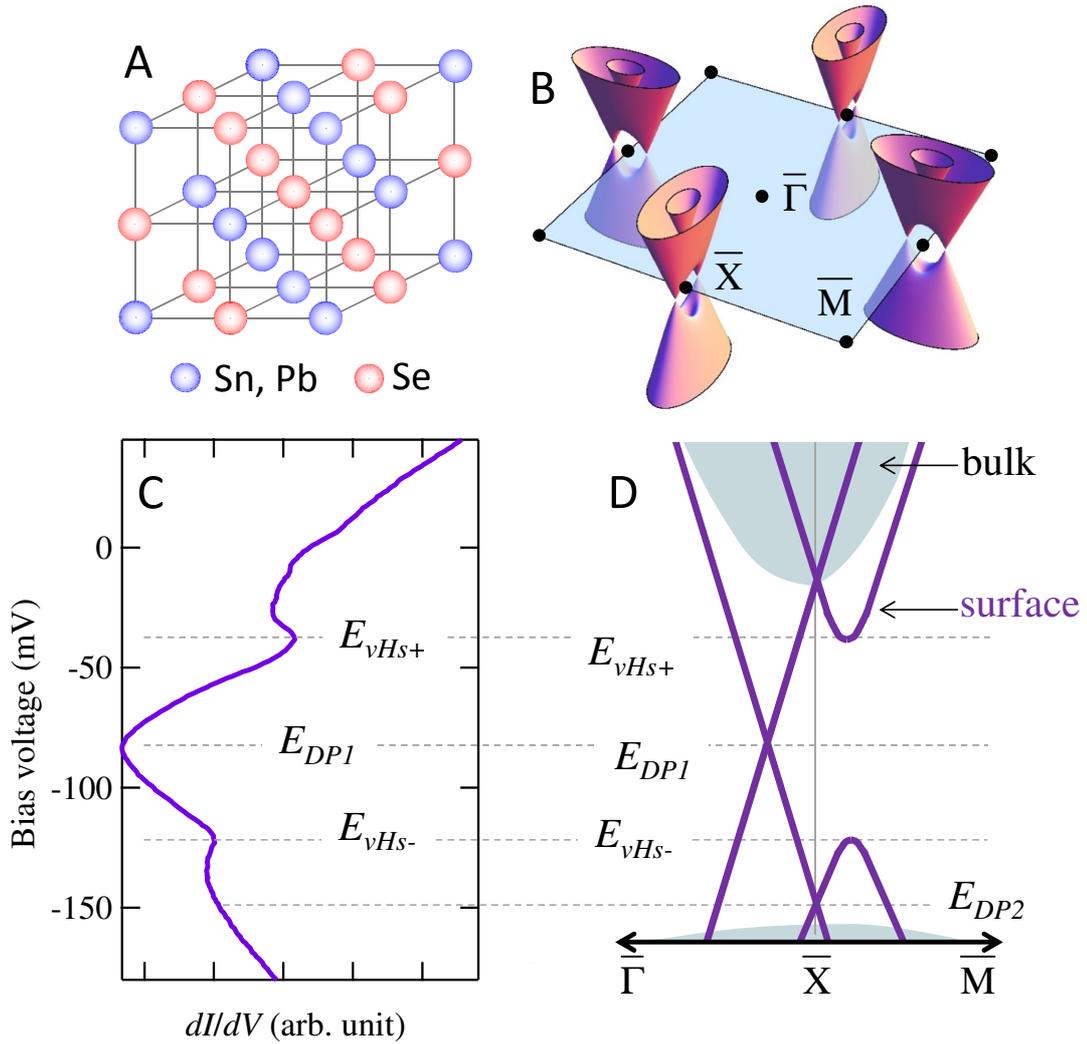

**Fig. 1.** Crystal structure and schematic band structure of TCIs. (**A**) Schematic crystal structure of $Pb_{1-x}Sn_xSe$. The top view reflects the [001] plane, which is the surface seen in STM. The Sn and Pb atoms are expected to be randomly distributed at the blue sites. (**B**) Schematic band structure of the surface state showing the surface Brillouin zone as a blue plane and the four pairs of Dirac nodes, each centered at the X point in momentum space. (**C**) Typical average dI/dV spectrum in zero field. (**D**) Schematic cuts along two high symmetry directions showing the surface state dispersion of one of the Dirac cones as well as the four important energy scales $E_{DP1}$, $E_{vH+}$, $E_{vHs-}$, and $E_{DP2}$ representing the Dirac point associated with the primary Dirac node, the two Van Hove singularities associated with the saddle points in the dispersion, and the Dirac point energy associated with the secondary Dirac node respectively.

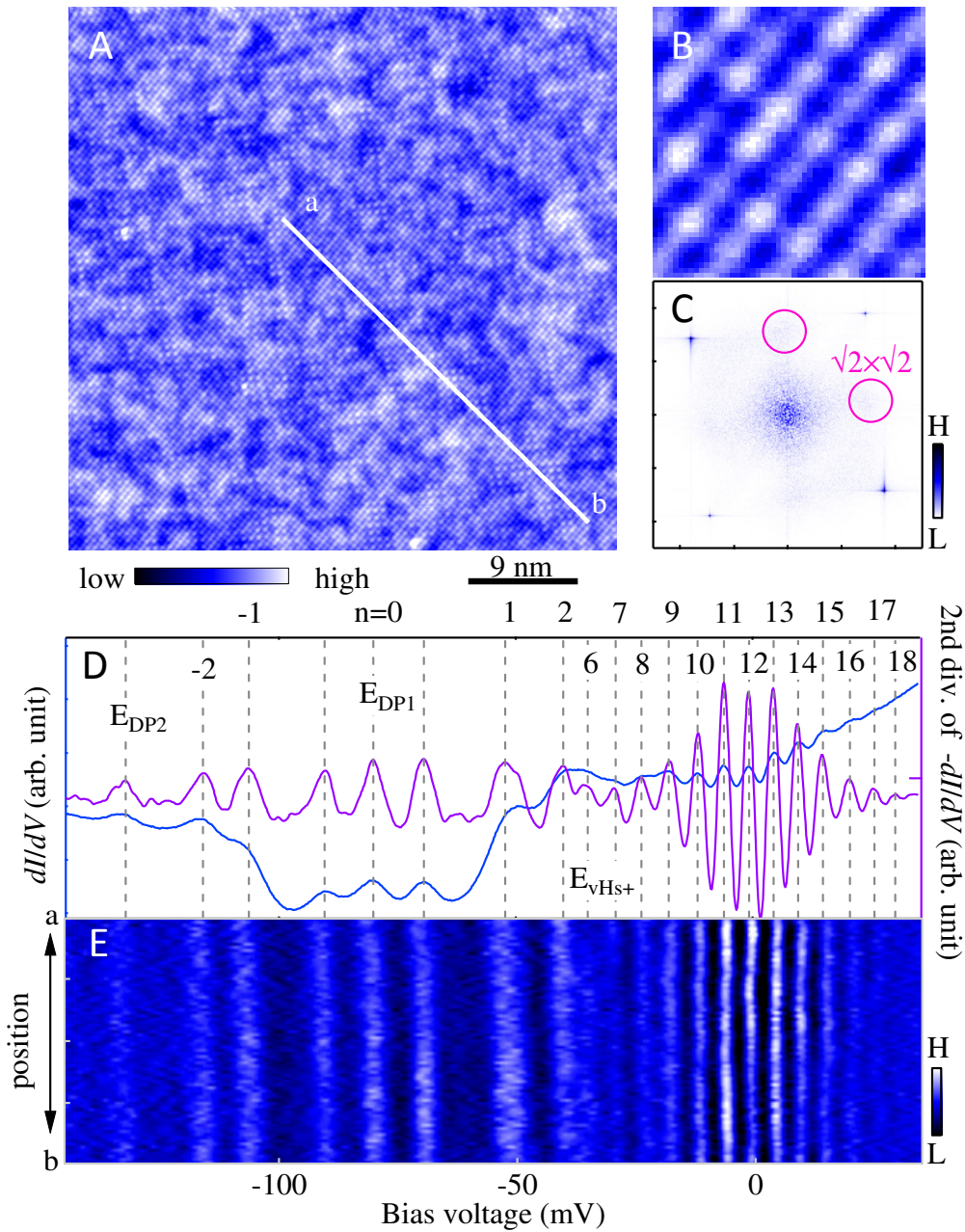

**Fig. 2.** STM image and position dependent spectra on $Pb_{1-x}Sn_xSe$. (**A**) Typical 40nm STM topographic image with clearly resolved atoms as seen in the smaller scale image in (**B**) and the Fourier transform in (**C**). (**D**) Averaged STM spectrum at 7.5T showing Landau levels (blue). The second derivative spectrum (purple) shows the LL peaks much more clearly. Second derivative of spectra were therefore used to identify the peak positions. (**E**) Image plot of the second derivative LL spectra at 7.5T as a function of position along the line shown in (**A**). The homogeneous nature of the sample is reflected in the lack of variation in the peak energy with position.

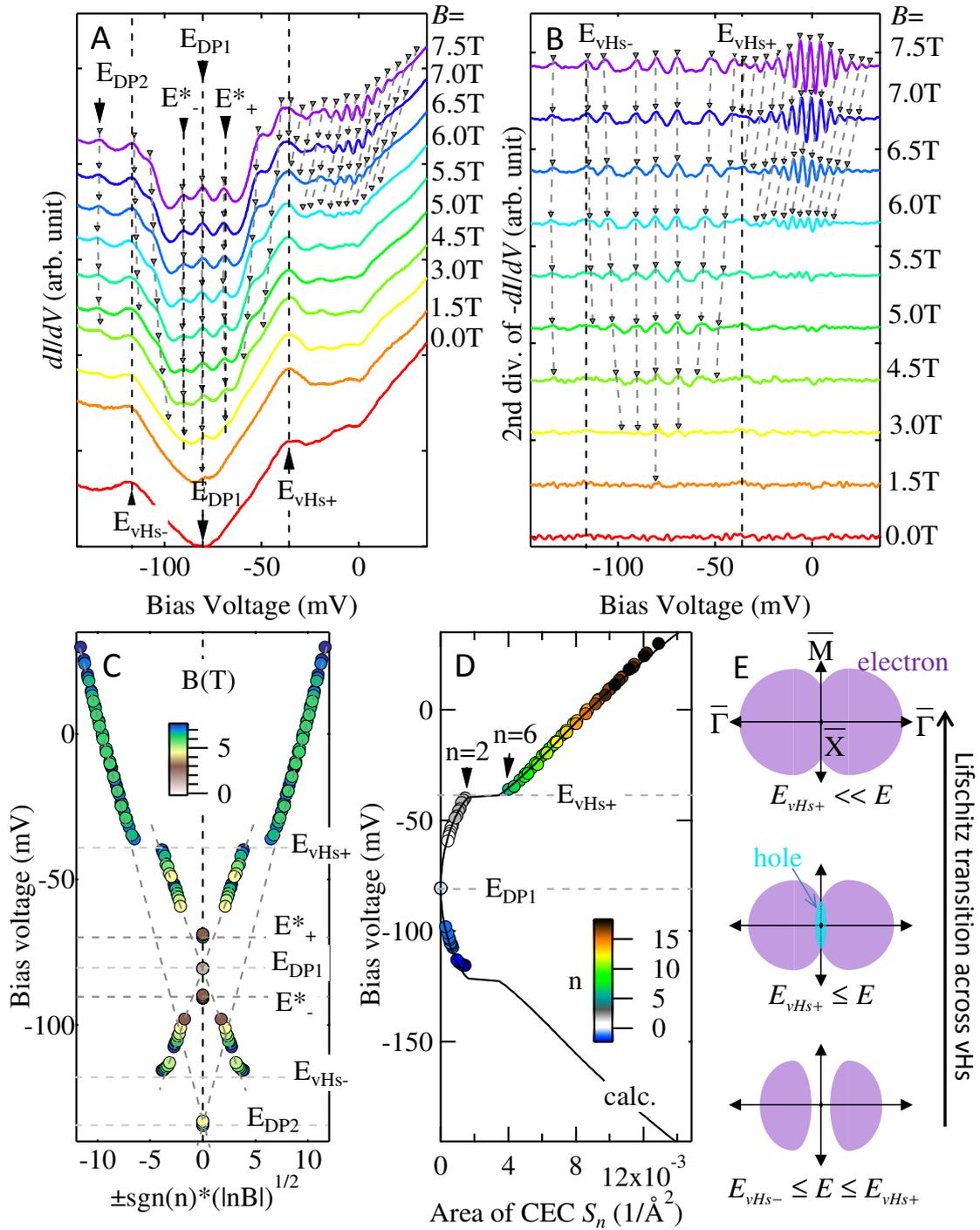

**Fig. 3.** VHS associate with change in Fermi surface topology measured by Landau level spectroscopy. (**A**) Linecut averaged STM spectra with increasing magnetic field. (**B**) Second derivative spectra that were used to trace the peak positions. (**C**) Plot of LL peak positions with sqrt nB. (**D**) Plot of theoretically calculated Fermi surface area with energy, overlaid with experimental LL peak positions as a function of nB (see supplemental information for details of calculation). (**E**) Schematic of evolution of the constant energy contours in momentum space with energy showing the Lifschitz transition.

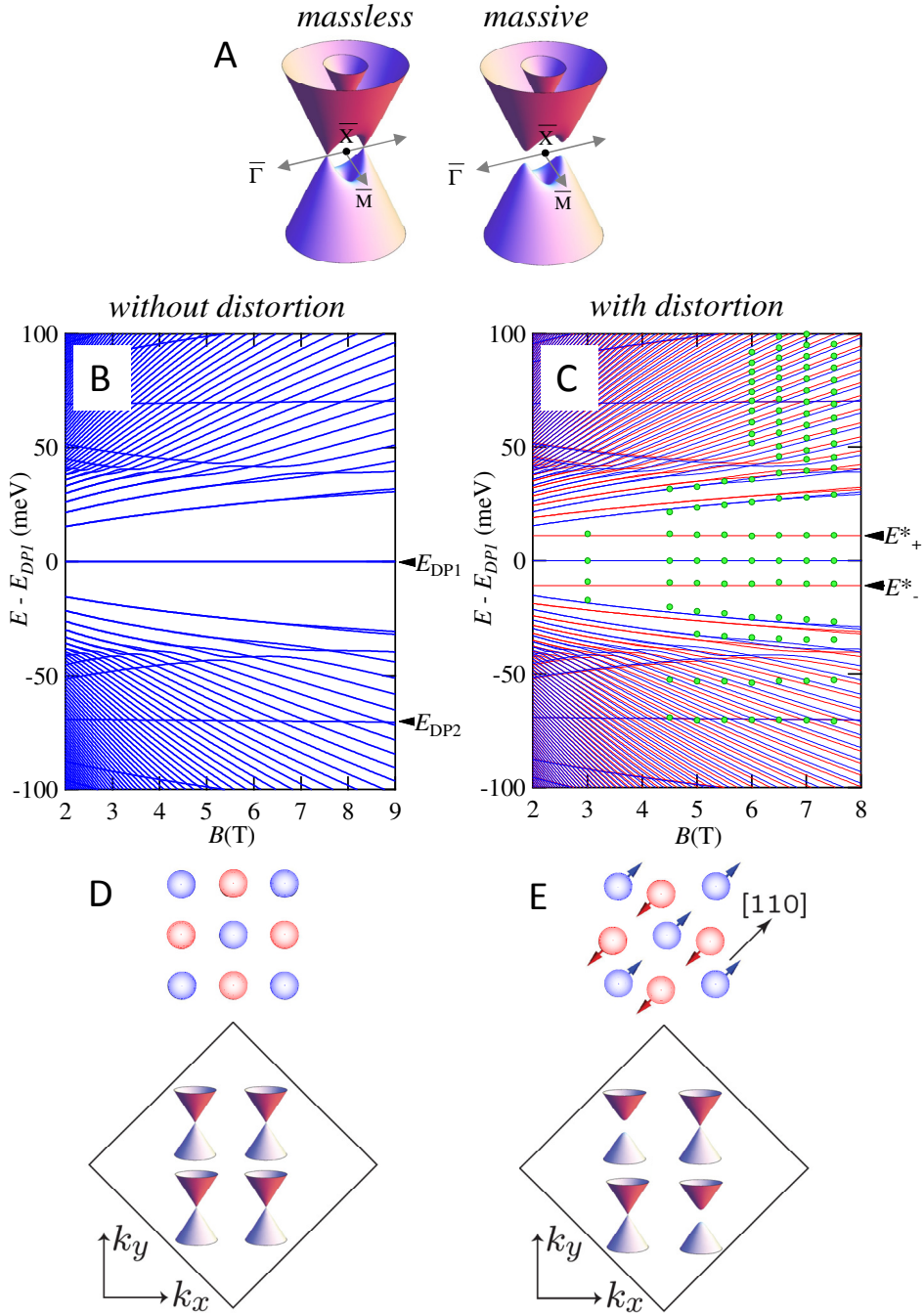

**Fig. 4.** Coexistence of massless and massive Dirac Fermions. Schematic band structure (**A**) and theoretical calculation of LL fan diagram without, (**B, D**) and with, (**C, E**) a symmetry breaking term added to the Hamiltonian, respectively. (**C**) shows comparison with the data points obtained from experimental LL spectra with the theoretical band structure parameters adjusted to match the data. The red and blue lines refer to the LLs from the distorted and undistorted surface states respectively. Schematic arrangement of surface atoms and band structure without (**D**), and with (**E**), a crystal distortion that leads to one broken mirror symmetry plane.